\documentclass[aps, pra, reprint,superscriptaddress]{revtex4-1}
\usepackage{graphicx}
\usepackage{dcolumn}
\usepackage{bm}
\usepackage{amsfonts}
\usepackage{amsmath}
\usepackage{amssymb}
\usepackage{color}

\usepackage[colorlinks=true,citecolor=blue]{hyperref}
\hypersetup{colorlinks=true,citecolor=blue,linkcolor=red,urlcolor=blue}

\def\be{\begin{equation}}
\def\ee{\end{equation}}

\def\kv{{\bf k}}
\def\qv{{\bf q}}

\def\Vc{{\cal V}}
\def\sigmav{{\bf \sigma}}

\newcommand{\h}[1]{{\hat {#1}}}
\newcommand{\hdg}[1]{{\hat {#1}^\dagger}}

\begin{document}

\title{Coupled spin-charge dynamics in helical Fermi liquids beyond the random phase approximation}
\author{Moslem Mir}
\affiliation{Department of Physics, Institute for Advanced Studies in Basic Sciences (IASBS), Zanjan 45137-66731, Iran}
\affiliation{Faculty of Science, University of Zabol (UOZ), Zabol 98615-538, Iran}
\author{Saeed H. Abedinpour}
\email{abedipour@iasbs.ac.ir}
\affiliation{Department of Physics, Institute for Advanced Studies in Basic Sciences (IASBS), Zanjan 45137-66731, Iran}
\affiliation{Research Center for Basic Sciences \& Modern Technologies (RBST),  Institute for Advanced Studies in Basic Sciences (IASBS), Zanjan 45137-66731, Iran}
\date{\today}

\begin{abstract}
We consider a helical system of fermions with a generic spin (or pseudospin) orbit coupling. 
Using the equation of motion approach for the single-particle distribution functions, and a mean-field decoupling of the higher order distribution functions, we find a closed form for the charge and spin density fluctuations in terms of the charge and spin density linear response functions. 
Approximating the nonlocal exchange term with a Hubbard-like local-field factor, we obtain coupled spin and charge density response matrix beyond the random phase approximation, whose poles give the dispersion of four collective spin-charge modes. 
We apply our generic technique to the well-explored two-dimensional system with Rashba spin-orbit coupling and illustrate how it gives results for the collective modes, Drude weight, and spin-Hall conductivity which are in very good agreement with the results obtained from other more sophisticated approaches. 
\end{abstract}
\maketitle
 
\section{introduction}\label{sec:intro}
Materials with strong spin-orbit couplings (SOC) have attracted an enormous interest recently not only because of their envisioned applications in spintronics~\cite{Zutic, Awshalom2009, Manchon2015} but also due to the eminent role they play in several fundamentally interesting phenomena such as topological phases of matter~\cite{ZHassan2010}, Majorana fermions~\cite{LiangFu2008}, etc. 
Coupling between a pseudospin or valley degree of freedom with the orbital motion of electrons in two-dimensional materials like graphene or monolayers of transition metal dichalcogenides have attracted a lot of interest too. Generation of synthetic gauge fields for ultracold gases~\cite{dalibard_rmp2011,goldman_rpp2014}, photonic~\cite{YPlotnik2016}, or mechanical~\cite{abbaszadeh_arxiv2016} systems are being actively explored both theoretically and experimentally. 

While several interesting single particle phenomena are associated with the helicity in all these systems, introduction of interparticle interactions could lead to unexpected properties~\cite{Ghchen1999-1,Ghchen1999-2, Magarill2001, Saraga2005, Dimitrova2005, Wang2005,Pletyukhov2006,Schliemann2006,Chesi2007,Badalyan2009,Ambrosetti2009,Nechaev2010, Badalyan2010,Zak2010,Chesi2011-1,Chesi2011-2,Ashrafi2012,Maiti2015-2}. 
In particular, it is well known that the breakdown of the Galilean invariance (GI) makes several quantities such as the Drude weight, optical response, and spin-Hall conductivity susceptible to many-body effects~\cite{Shekhter,Farid2006,Amit2011,Abedinpour2011,Maiti2015}. 
For the collective modes, going beyond the standard random phase approximation (RPA) is necessary in order to obtain the correct dispersion even in the long wavelength limit for broken Galilean invariant systems~\cite{Abedinpour2011}.

In a conventional electron liquid, the dispersion of collective charge i.e., \textit{plasmon} mode could be obtained from the interacting linear density-density response function, which within the celebrated RPA reads~\cite{GV_book}
\be
\chi^{\rm RPA}(\qv,\omega)=\left[1- \chi^0(\qv,\omega)v(q)\right]^{-1} \chi^0(\qv,\omega),
\ee
where $\chi^0(\qv,\omega)$ is the noninteracting density-density response function and $v(q)$ is the Fourier transform of the particle-particle interaction potential. Improvements upon the RPA could be generally thought in two directions: i) replacing the non-interacting density response function with the proper one and including the lowest order irreducible diagrams in it~\cite{GV_book} as is done \textit{e.g.}, in time-dependent Hartree-Fock (TDHF) approach, or ii) replacing the bare interparticle interaction $v(q)$ with an effective one $w(q,\omega)$ through the introduction of the many-body local field factors (LFF) which in its simplest form could be written as $w(q,\omega)=v(q)[1-G(q,\omega)]$. The first approach is usually very complicated and hard to go beyond the lowest order corrections. On the other hand, historically the concept of many-body LFFs have received a lot of interest and several, mainly self-consistent formulations for its evaluation are available~\cite{GV_book} for conventional electron liquids i.e., in the absence of SOC.

While there are several attempts to go beyond the RPA for helical liquids, in particular for electronic systems with SOC~\cite{Amit2011,Maiti2015} and graphene~\cite{Abedinpour2011}, to the best of our knowledge the concept of LFFs has never been extended to these systems. Our aim in this work is to follow the steps which lead to the first version of the LFF for electron liquids by John Hubbard~\cite{Hubbard_1957}, for a Fermi liquid with a generic spin or pseudospin orbit coupling. In this way, we find a closed expression for the interacting spin-density response functions in terms of the non-interacting response functions and the Hubbard LFF. The singularities of these responses provide the dispersions of the coupled spin-charge collective modes. Then, we apply this formalism to a two-dimensional (2D) system with Rashba SOC, whose many-body properties has been extensively explored in the past few years with different techniques. We find that our results for the long wavelength behavior of the collective modes, for the many-body modification of the Drude weight, and for the spin-Hall conductivity agrees well with the findings of others. 

The rest of this paper is organized as follows. In Sec.~\ref{sect:model} we introduce our generic Hamiltonian and the basic notations. In Sec.~\ref{sect:eom} we find the linear spin-density response functions beyond the RPA using the equation of motion method and the mean field approximation. The results of this section have been used to find the collective modes, Drude weight and spin-Hall conductivity of a 2D Rashba system in Sec.~\ref{sect:rashba2deg}. In Sec.~\ref{sect:summary} we summarize our main findings and discuss the future directions for further development of the LFFs for helical systems. Finally, we have dedicated two appendices to present the details of our equation of motion and mean-field decoupling technique (Appendix~\ref{app:eom}) as well as the spin-density response functions of a 2D Rashba system (Appendix~\ref{app:rashba}).
 
\section{Model and basic notations}\label{sect:model}
The Hamiltonian of an interacting D-dimensional 2-band Fermi system with  a generic spin-orbit coupling (SOC) and a spin-dependent external potential can be written as
\be\label{hamil}
\begin{split}
\h{H}&=\h{T}+\h{V}_{\rm ext} +\h{V}_{\rm int} \\
&=
\sum_{\kv,\sigma,\sigma'}  \varepsilon_{\sigma,\sigma'}(\kv) \hdg{a}_{\kv,\sigma}  \h{a}_{\kv,\sigma'} 
+ \frac{1}{\Vc}  \sum_{i,\qv} \phi_{\rm ext}^i(-\qv)  S^{i}_{\qv} \\
&+\frac{1}{2 \Vc}  \sum_{\qv} v(q)  :S^{0}_{\qv} S^{0}_{-\qv}: .
\end{split}
\ee
Here, the single-particle energy matrix $\varepsilon(\kv)$ is naturally non-diagonal in the presence of SOC,
 $\hdg{a}_{\kv,\sigma}$ and $\h{a}_{\kv,\sigma}$ respectively create and annihilate an electron with wave vector $\kv$ and spin $\sigma $, ${\cal V}$ is the D-dimensional volume of system,  $i$ runs over the particle density ($i=0$) and 3 cartesian components of spin ($i=1, 2, 3$), $\phi_{\rm ext}^i(-\qv)$ is the external potential coupled to the $i$-th component of the spin-density, $v(q)$ is the D-dimensional Fourier transform of the interparticle interaction potential,
$:{\h A}{\h B} \cdots :$ enforces the normal ordering of the enclosed creation and annihilation operators, and finally
the spin-density operators are defined as
\be
\hat{S}^{i}_\qv=\sum_{\kv,\sigma\sigma'}\hdg{a}_{\kv-\qv,\sigma}\tau^i_{\sigma,\sigma'}\h{a}_{\kv,\sigma'},
\ee
where $\tau^0=\mathbb{I}$, and $\tau^{i\neq 0}$ refer to the standard $2\times2$ Pauli matrices. 
Note that the $q=0$ term in the last term of Eq.~\eqref{hamil} represents the Hartree contribution to the interacting energy and should be discarded for the jellium model of electron liquids~\cite{GV_book}.
With the help of a suitable unitary transformation $U(\kv)$, the single-particle energy matrix $\varepsilon(\kv)$ could be diagonalized to give two energy bands $\varepsilon_{\kv,\pm}$, and the Hamiltonian~\eqref{hamil} in the band basis reads
\be\label{eq:hamil_band}
\begin{split}
\h{H}&=\sum_{\kv,\mu} \varepsilon_{\kv,\mu}  \hdg{c}_{\kv,\mu} \h{c}_{\kv,\mu} 
+ \frac{1}{\Vc}  \sum_{i,\qv} \phi_{\rm ext}^i(-\qv)  S^{i}_{\qv} \\
&+\frac{1}{2\Vc}  \sum_{\qv} v(q)  :S^{0}_{\qv} S^{0}_{-\qv}:
 .
\end{split}
\ee
Here, $\hdg{c}_{\kv,\mu}$ and $\h{c}_{\kv,\mu}$ are creation and annihilation operators of electrons in the band basis, and the spin-density operators in the new basis read
\be\label{spin_density}
\h{S}^{i}_\qv=\sum_{\kv,\mu\nu} F^{i}_{\mu,\nu}(\kv-\qv;\kv)\hdg{c}_{\kv-\qv,\mu}\h{c}_{\kv,\nu},
\ee
with the form factor matrices being defined as
\be\label{eq:formfactor}
F^{i}(\kv-\qv;\kv) = U^\dagger(\kv-\qv) \tau^i U(\kv),
\ee
where the matrix product is understood in the right hand side (RHS) of Eq.~(\ref{eq:formfactor}), and clearly we have $F^{i}_{\mu,\nu}(\kv-\qv;\kv)=F^{i*}_{\nu,\mu}(\kv;\kv-\qv)$. The explicit expressions for these form factors depend on the specific form of the spin-orbit coupling of interest, and for the special case of Rashba SOC they are provided in Sec. \ref{sect:rashba2deg}.

\section{Equation of motion}\label{sect:eom}
In order to obtain the spin-density linear response functions, we write the Heisenberg equation of motion for the single particle distribution function
$\langle \hdg{c}_{\kv-\qv,\mu} \h{c}_{\kv,\nu} \rangle$~\cite{niklasson_prb74,solyom_v3,mahan_book,Simion_thesis} ($\hbar=1$)
\be\label{eq:eom}
i \frac{d}{dt}\left\langle \hdg{c}_{\kv-\qv,\mu} \h{c}_{\kv,\nu}\right\rangle
=\left\langle \left[\hdg{c}_{\kv-\qv,\mu} \h{c}_{\kv,\nu},\h{H}\right]\right\rangle. 
\ee
Note that in the presence of an external perturbation $\phi_{\rm ext}$, we have 
$\left\langle \hdg{c}_{\kv',\mu} \h{c}_{\kv,\nu}\right\rangle =\delta_{\kv,\kv'}\delta_{\mu,\nu} n_{\kv,\mu}+{\cal O}(\phi_{\rm ext})$, where $n_{\kv,\mu}$ is the equilibrium part of the occupation number of fermions.
With some straightforward algebra (see, Appendix~\ref{app:eom} for the details), for the first two terms of the Hamiltonian~\eqref{eq:hamil_band} we find~\cite{foot:invariance} 
\be\label{eq:cc-T}
\left\langle \left[\hdg{c}_{\kv-\qv,\mu} \h{c}_{\kv,\nu},\h{T}\right]\right\rangle =
\left(\varepsilon_{\kv,\nu}-\varepsilon_{\kv-\qv,\mu}\right)\left\langle \hdg{c}_{\kv-\qv,\mu} \h{c}_{\kv,\nu}\right\rangle,
\ee
and
\be\label{eq:cc-Vext}
\begin{split}
\left\langle\left[\hdg{c}_{\kv-\qv,\mu} \h{c}_{\kv,\nu},\h{V}_{\rm ext}\right]\right\rangle &
\approx \frac{1}{\Vc}\left(n_{\kv-\qv,\mu}-n_{\kv,\nu}\right) \\
&\times \sum_j \phi^j_{\rm ext}(\qv) F^j_{\nu,\mu}(\kv;\kv-\qv).
\end{split}
\ee
On the RHS of Eq.~(\ref{eq:cc-Vext}), only terms to linear order in the external potential are retained, therefore the expectation values are safely evaluated with respect to the equilibrium (i.e., unperturbed) state.

For the interaction term of the Hamiltonian~\eqref{eq:hamil_band}, after some lengthly but straightforward algebra find
\begin{widetext}
\be\label{eq:cc-P1}
\begin{split}
&\left\langle\left[\hdg{c}_{\kv-\qv,\mu} \h{c}_{\kv,\nu},\hat{V}_{\rm int}\right]\right\rangle
=\frac{1}{\Vc} \sum_{\qv'}v(\qv')\sum_{\kv',\mu',\nu'}
F^{0}_{\mu',\nu'}(\kv'-\qv';\kv')\\
&\times\sum_\gamma\left[F^{0}_{\nu,\gamma}(\kv;\kv-\qv')
\left\langle\hdg{c}_{\kv-\qv,\mu} \hdg{c}_{\kv'-\qv',\mu'}\h{c}_{\kv',\nu'}\h{c}_{\kv-\qv',\gamma}\right\rangle
-F^{0}_{\gamma,\mu}(\kv-\qv+\qv';\kv-\qv)\left\langle\hdg{c}_{\kv-\qv+\qv',\gamma} \hdg{c}_{\kv'-\qv',\mu'}\h{c}_{\kv',\nu'}\h{c}_{\kv,\nu}\right\rangle\right].
\end{split}
\ee
\end{widetext}
Now, using the mean-field decoupling of the quartic terms in creation and annihilation operators into quadratic terms 
$
\langle\hdg{c}_1\hdg{c}_2\h{c}_3\h{c}_4 \rangle\approx  \langle \hdg{c}_1\h{c}_4\rangle  \langle\hdg{c}_2 \h{c}_3\rangle
- \langle \hdg{c}_1\h{c}_3\rangle \langle \hdg{c}_2 \h{c}_4\rangle
$,
then keeping up to linear terms in the external potentials and also discarding the self-energy contribution to the single particle energies (see, Appendix~\ref{app:eom} for the details), we obtain
\begin{widetext}
\be\label{eq:cc-P}
\begin{split}
\left \langle  \left[\hdg{c}_{\kv-\qv,\mu} \h{c}_{\kv,\nu},\hat{V}_{\rm int}\right]\right\rangle=
\frac{1}{\Vc}\left(n_{\kv-\qv,\mu}-n_{\kv,\nu}\right)  
&  \Bigg[
v(q) F^{0}_{\nu,\mu}(\kv;\kv-\qv)
\langle\h{S}^0_\qv \rangle    \\
&- \sum_{\kv',\mu',\nu'} v(\kv-\kv')
 F^{0}_{\nu,\nu'}(\kv;\kv') F^{0}_{\mu',\mu}(\kv'-\qv;\kv-\qv)\left\langle \hdg{c}_{\kv'-\qv,\mu'} \h{c}_{\kv',\nu'} \right\rangle\Bigg].
\end{split}
\ee
\end{widetext}
Note that the first line on the RHS of the above expression represents the standard random-phase approximation, while the second line is the effects of exchange beyond the RPA. 

Now, using the completeness relation for the Pauli matrices
$\sum_{i=0}^3 \tau^i_{\alpha,\beta}\tau^i_{\gamma\delta}=2 \delta_{\alpha,\delta}\delta_{\gamma\beta}$, 
 we find the following very useful general identity for the form factors
\be\label{eq:decoupling-form-factor}
\begin{split}
 F^{0}_{\nu,\nu'}&(\kv;\kv')  F^0_{\mu',\mu}(\kv'-\qv;\kv-\qv)
\\&=\frac{1}{2}\sum_{i=0}^3
 F^{i}_{\nu,\mu}(\kv;\kv-\qv) F^{i}_{\mu',\nu'}(\kv'-\qv;\kv'),
 \end{split}
\ee
which is independent of the specific form of form factors. Now, using Eq.~\eqref{eq:decoupling-form-factor} in the second line of Eq.~(\ref{eq:cc-P}),  
summing all the contributions to the RHS of Eq.~(\ref{eq:eom}) and then taking Fourier transformation with respect to time, we find
\be\label{eq:cc_full}
\begin{split}
&\langle\hdg{c}_{\kv-\qv,\mu} \h{c}_{\kv,\nu}\rangle=
\\
&\frac{1}{\Vc}\frac{n_{\kv-\qv,\mu}-n_{\kv,\nu}}{\omega +\varepsilon_{\kv-\qv,\mu}-\varepsilon_{\kv,\nu}}  \sum_j F^j_{\nu,\mu}(\kv;\kv-\qv) \phi^j_{\rm ext}(\qv)
 \\
&+v(q) \frac{1}{\Vc}\frac{n_{\kv-\qv,\mu}-n_{\kv,\nu}}{\omega +\varepsilon_{\kv-\qv,\mu}-\varepsilon_{\kv,\nu}}   F^{0}_{\nu,\mu}(\kv;\kv-\qv)\langle S^0_\qv \rangle
 \\
&-\frac{1}{2\Vc}\sum_{j=0}^3 F^{j}_{\nu,\mu}(\kv;\kv-\qv)
\frac{n_{\kv-\qv,\mu}-n_{\kv,\nu}}{\omega +\varepsilon_{\kv-\qv,\mu}-\varepsilon_{\kv,\nu}}
\\
&\times \sum_{\kv',\mu',\nu'} v(\kv-\kv')  F^{j}_{\mu',\nu'}(\kv'-\qv;\kv')  \langle\hdg{c}_{\kv'-\qv,\mu'} \h{c}_{\kv',\nu'}\rangle.
\end{split}
\end{equation}
The non-local form of the last line on the RHS of Eq.~(\ref{eq:cc_full}) makes it impossible to find a closed form for the density fluctuations. Following Hubbard~\cite{Hubbard_1957,GV_book,mahan_book}, we approximate $v(\kv-\kv') \approx v_{\rm H}(q)$ in the non-local term of Eq.~\eqref{eq:cc_full}. This is equivalent to the summation of ladder diagrams with screened interaction and $v_{\rm H}(q)=v(\sqrt{k_{\rm F}^2+q^2})$ is usually adopted for electron liquids~\cite{solyom_v3, GV_book}, where $k_{\rm F}$ is the Fermi wave vector.
Furthermore, multiplying both sides of Eq.~\eqref{eq:cc_full} by $F^{i}_{\mu,\nu}(\kv-\qv;\kv)$, and summing over $\kv$, $\mu$, and $\nu$ we obtain
\begin{equation}\label{RPA_H}
\begin{split}
\langle S^{i}_{\qv}\rangle&=
\sum_j\chi^{0}_{i,j}(\qv,\omega)\phi^j_{\rm ext}(\qv)+
v(q)\chi^{0}_{i,0}(\qv,\omega) \langle S^{0}_{\qv} \rangle
\\
&-\frac{1}{2}v_{\rm H}(q)\sum_j \chi^{0}_{i,j}(\qv,\omega) \langle S^j_\qv \rangle.
\end{split}
\end{equation}
Here, the non-interacting spin-density response functions are defined as
\be\label{eq:chi0_matrix}
\begin{split}
&\chi^{0}_{i,j}(\qv,\omega)=\langle\langle  S^{i}_{\qv}; S^{j}_{-\qv}\rangle\rangle_\omega 
\\
&=\frac{1}{\Vc}\sum_{\kv,\mu,\nu} 
\frac{n_{\kv-\qv,\mu}-n_{\kv,\nu}}{\omega +\varepsilon_{\kv-\qv,\mu}-\varepsilon_{\kv,\nu}}
F^{i}_{\mu,\nu}(\kv-\qv;\kv)F^{j}_{\nu,\mu}(\kv;\kv-\qv).
\end{split}
\ee
With some straightforward rearrangements, we can rewrite Eq.~\eqref{RPA_H} as
\be\label{eq:s_chi_phi}
\begin{split}
{\vec S}_\qv&=\chi(\qv,\omega) {\vec \phi}_{\rm ext}(\qv)\\
&=\left[\mathbb{I}-\chi^0(\qv,\omega)W(q)\right]^{-1}\chi^0(\qv,\omega) {\vec \phi}_{\rm ext}(\qv),
\end{split}
\ee
where $\chi(\qv,\omega)$ is the $4\times 4$ interacting spin-density response matrix, a 4D vector ${\vec A}$ is defined as $\left(A^0, A^1,A^2,A^3\right)^T$, and  the elements of the effective interaction matrix read
\be\label{eq:effective-interaction}
\begin{split}
W_{i,j}(q)&=v(q)\delta_{i,0}\delta_{j,0}-\frac{1}{2}v_{\rm H}(q)\delta_{i,j}\\
&=v(q)\left[\delta_{i,0}\delta_{j,0}-G_{\rm H}(q)\delta_{i,j}\right],
\end{split}
\ee
where the Hubbard local field factor in the second line is defined as $G_{\rm H}(q)=v_{\rm H}(q)/[2v(q)]$. Eqs.~\eqref{eq:s_chi_phi} and~\eqref{eq:effective-interaction}, which are independent of the specific form of the SOC, compromise the main general results of the present paper.

Dispersions of the collective spin-charge modes could be obtained from the solutions of 
\be\label{det}
\det \left[\mathbb{I}-\chi^0(\qv,\omega)W(q)\right]=0.
\ee
We recall that within the RPA, one simply ignores the non-local exchange term in Eq.~(\ref{eq:cc_full}), which is equivalent to putting $v_{\rm H}(q)=0$ in Eq.~(\ref{RPA_H}), and finds
$1-v(q)\chi^0_{0,0}(\qv,\omega)=0$ for the dispersion of collective modes.
Therefore, within the RPA, one simply finds one charge mode and no spin modes. 
Before moving to apply our formalism to the well-explored two-dimensional system with Rashba spin-orbit coupling, let us briefly comment on the case of ultrashort range interactions.

\subsection{Ultrashort range interactions\label{sec:shortrange}}
If the interaction between particles is ultrashort $v(r)=U\delta(r)$ \textit{e.g.}, in the case of neutral ultracold atomic gases with repulsive $s$-wave interactions, as the Fourier transform of the interaction becomes constant $v(q)=U$, the exchange term in Eq.~\eqref{eq:cc_full} becomes local and without any further approximation we arrive at Eq.~\eqref{RPA_H} with $v(q)=v_{\rm H}(q)=U$. 
The effective interaction matrix becomes $W=(U/2)\, \textrm{diag}\{1,-1,-1,-1\}$, and one immediately realizes that the effective interaction between identical spins, as the Pauli exclusion principle also implies, correctly vanishes~\cite{Zhang_pra2013}. 

\section{Rashba spin-orbit coupling}\label{sect:rashba2deg}
In this section we apply the formalism developed in previous section to a 2D system with Rashba SOC.  The single-particle energy matrix of this system is (again, $\hbar=1$)
\begin{equation}\label{single}
\varepsilon(\kv)=\frac{k ^2}{2m} \mathbb{I}+\alpha \left (\kv \times \sigmav \right)\cdot \h{z},
\end{equation}
where $m$ is the particle mass, $\alpha$ is the strength of Rashba SOC, and $\h{z}$ is the unit vector in the direction perpendicular to the 2D plane. This matrix could be easily diagonalized to give the energy dispersions
\be\label{e_k}
\varepsilon_{k,\pm}=\frac{ k^2}{2m}\pm \alpha  |k|,
\ee
and the unitary transformation matrix which diagonalizes the Rashba single-particle energy matrix could be written as
\be\label{eq:untary}
U(\kv)=\frac{1}{\sqrt{2}}\left(\begin{array}{cc} 
1& 1\\ i e^{i\varphi_{k}}& -i e^{i\varphi_{k}}
\end{array}\right),
\ee
where $\varphi_k$ is the angle between $\kv$ and the $x$-axis.
The different elements of the form factors could be obtained from Eq.~\eqref{eq:formfactor}, and read
\be\label{eq:formfactors}
\begin{split}
&F^{0}_{\mu,\nu}(\kv-\qv;\kv)=\frac{1}{2}\left(1+\mu \nu e^{i(\varphi_{k}-\varphi_{k-q})}\right),\\
&F^{1}_{\mu,\nu}(\kv-\qv;\kv)=\frac{i}{2}\left(\nu  e^{i\varphi_{k}} -\mu e^{-i\varphi_{k-q}}\right),\\
&F^{2}_{\mu,\nu}(\kv-\qv;\kv)=\frac{1}{2}\left(\nu  e^{i\varphi_{k}} +\mu e^{-i\varphi_{k-q}}\right),\\
&F^{3}_{\mu,\nu}(\kv-\qv;\kv)=\frac{1}{2}\left(1-\mu \nu e^{i(\varphi_{k}-\varphi_{k-q})}\right).
\end{split}
\ee
Now, if we choose $\qv= q \h{e}_x$, it could be easily shown~\cite{Maiti2015} that only 6 independent elements of $\chi^0(\qv,\omega)$ are nonzero: 
$\chi^{0}_{0,0}$, $\chi^{0}_{1,1}$, $\chi^{0}_{2,2}$, $\chi^{0}_{3,3}$, $\chi^{0}_{0,2}=\chi^{0}_{2,0}$, and $\chi^{0}_{1,3}=-\chi^{0}_{3,1}$ (see, Appendix~\ref{app:rashba} for more details).
Then, the secular equation~(\ref{det}) results in two decoupled set of equations as
\be\label{mode1} 
\begin{split}
&\left[1-[v(q)-\frac{1}{2}v_{\rm H}(q)]\chi^{0}_{0,0}(\qv,\omega)\right]
\left[1+\frac{1}{2}v_{\rm H}(q)\chi^{0}_{2,2}(\qv,\omega)\right]
\\
&+\frac{1}{2}v_{\rm H}(q)\left[v(q)-\frac{1}{2}v_{\rm H}(q)\right]\left[\chi^{0}_{0,2}(\qv,\omega)\right]^2 =0,
\end{split}
\ee
and
\be\label{mode2}
\begin{split}
&\left[1+\frac{1}{2}v_{\rm H}(q) \chi^{0}_{1,1}(\qv,\omega)\right]
\left[1+\frac{1}{2}v_{\rm H}(q) \chi^{0}_{3,3}(\qv,\omega)\right]
\\
&+\frac{v^2_{\rm H}(q)}{4}\left[\chi^{0}_{1,3}(\qv,\omega)\right]^2  =0.
\end{split}
\ee
From these equations it is clear that the longitudinal (i.e., $x$) and perpendicular (i.e., $z$) components of the spin-modes are coupled together, and the charge-mode is coupled to the transverse (i.e., $y$) component of the spin-mode. We recall that we have chosen $\qv=q\h{e}_x$ and the longitudinal and transverse directions are defined accordingly. As we will discuss in the following, this coupling between charge and transverse spin modes in particular is responsible for the interaction induced modifications of the Drude weight and spin-Hall conductivity in Rashba system.  

Below, we investigate the coupled spin-density modes of i) a two-dimensional system of ultracold gases with Rashba SOC and ultrashort interparticle interaction, and ii) a Rashba two-dimensional electron gas where the interparticle interaction is coulombic. Then, we discuss the effects of interaction on the spin-Hall conductivity of these systems.

\subsection{Collective modes of a 2D system of ultracold atoms with Rashba SOC}
In this subsection, we discuss the collective modes of a two-dimensional system of ultracold atomic gases with ultrashort particle-particle interactions and a spin-orbit coupling of Rashba form.
The sign and strength of the short range interaction $U$ in ultracold gases could be tuned through the Feshbach resonance~\cite{chin_rmp2010} and a synthetic spin-orbit coupling could be also induced \textit{e.g.}, with laser beams~\cite{dalibard_rmp2011,goldman_rpp2014}. The dispersions of 4 collective spin-density modes could be obtained after replacing $v(q)$ and $v_{\rm H}(q)$ in Eqs.~(\ref{mode1}) and~(\ref{mode2}) with $U$, which result in
\be\label{mode1_u} 
\left[1-\frac{U}{2}\chi^{0}_{0,0}(\qv,\omega)\right]\left[1+\frac{U}{2}\chi^{0}_{2,2}(\qv,\omega)\right]+\frac{U^2}{4}\left[\chi^{0}_{0,2}(\qv,\omega) \right]^2=0,
\ee
and
\be\label{mode2_u}
\left[1+\frac{U}{2}\chi^{0}_{1,1}(\qv,\omega)\right]\left[1+\frac{U}{2}\chi^{0}_{3,3}(\qv,\omega)\right]+\frac{U^2}{4}\left[\chi^{0}_{1,3}(\qv,\omega)\right]^2  =0.
\ee
We should note that in this case our results are identical to the ones of Zhang \textit{et al.}~\cite{Zhang_pra2013} obtained from generalized RPA. Solutions of Eqs.~\eqref{mode1_u} and \eqref{mode2_u} result in three massive modes and one acoustic mode. 
At $q=0$, only $\chi^0_{11}$, $\chi^0_{22}$, and $\chi^0_{33}$ are non-zero [see, Eq.~\eqref{eq:q=0} in Appendix~\ref{app:rashba}]. Therefore the above equations become decoupled and the masses of three gapped modes are obtained from
\be\label{eq:w11_33}
\begin{split}
\frac{2}{u}&=\eta+\frac{\omega}{8m \alpha^2}L(\omega), ~~~\textrm{for 11 and 22 modes},\\
\frac{1}{u}&=\eta+\frac{\omega}{8m \alpha^2}L(\omega), ~~~\textrm{for 33 mode}.
\end{split}
\ee
Here, $u=mU/(2\pi)$, $\eta= {\textrm min}[1,1/(2\lambda)]$,  where $\lambda=m^2\alpha^2/(2n\pi)$ is the dimensionless SOC strength, with $n$ being the particle density of system and
\be
L(\omega)=\ln\left[\frac{(\omega-\omega_{-}+i0^+)(\omega+\omega_{+}+i0^+)}{(\omega-\omega_{+}+i0^+)(\omega+\omega_{-}+i0^+)}\right],
\ee
with $\omega_{\pm}=2 m\alpha^2|\sqrt{1+\bar{\varepsilon}_{F}/\lambda}\mp 1|$, where $\bar{\varepsilon}_{F}=m\varepsilon_{F}/(n\pi)$ is the dimensionless Fermi energy.
Note that the Fermi energy $\varepsilon_{\rm F}$ should be larger than the bottom of the lower band $-m\alpha^2/2$ and it is easily shown that in the two-band regime (i.e., $\varepsilon_{\rm F}>0$) $\bar{\varepsilon}_{F}=1-2\lambda$ and in the single band regime (i.e., $\varepsilon_{\rm F}<0$) $\bar{\varepsilon}_{F}=1/(4\lambda)-\lambda$.
It is clear from Eq.~\eqref{eq:w11_33} that longitudinal and transverse modes are degenerate in the long wavelength limit while excitation of the perpendicular mode requires considerably lower energy. 
In the two band regime, $\eta=1$ and Eq.~\eqref{eq:w11_33} reduces to Eq.~(33) of Maiti \textit{et al.} in Ref.~\cite{Maiti2015} which are obtained for a 2DEG with Rashba SOC with diagrammatic techniques, while in the single band regime, we have $0<\eta<1$ and above results are obtained for the first time here. 
In the limiting cases of very weak or strong interaction strengths, it is possible to obtain analytic solutions for Eqs.~\eqref{eq:w11_33}. As these results are discussed in detail in Refs.~\cite{Maiti2015} and~\cite{Zhang_pra2013}, we do not repeat them here. 

\begin{figure}
\centering
\includegraphics[width=0.5\textwidth]{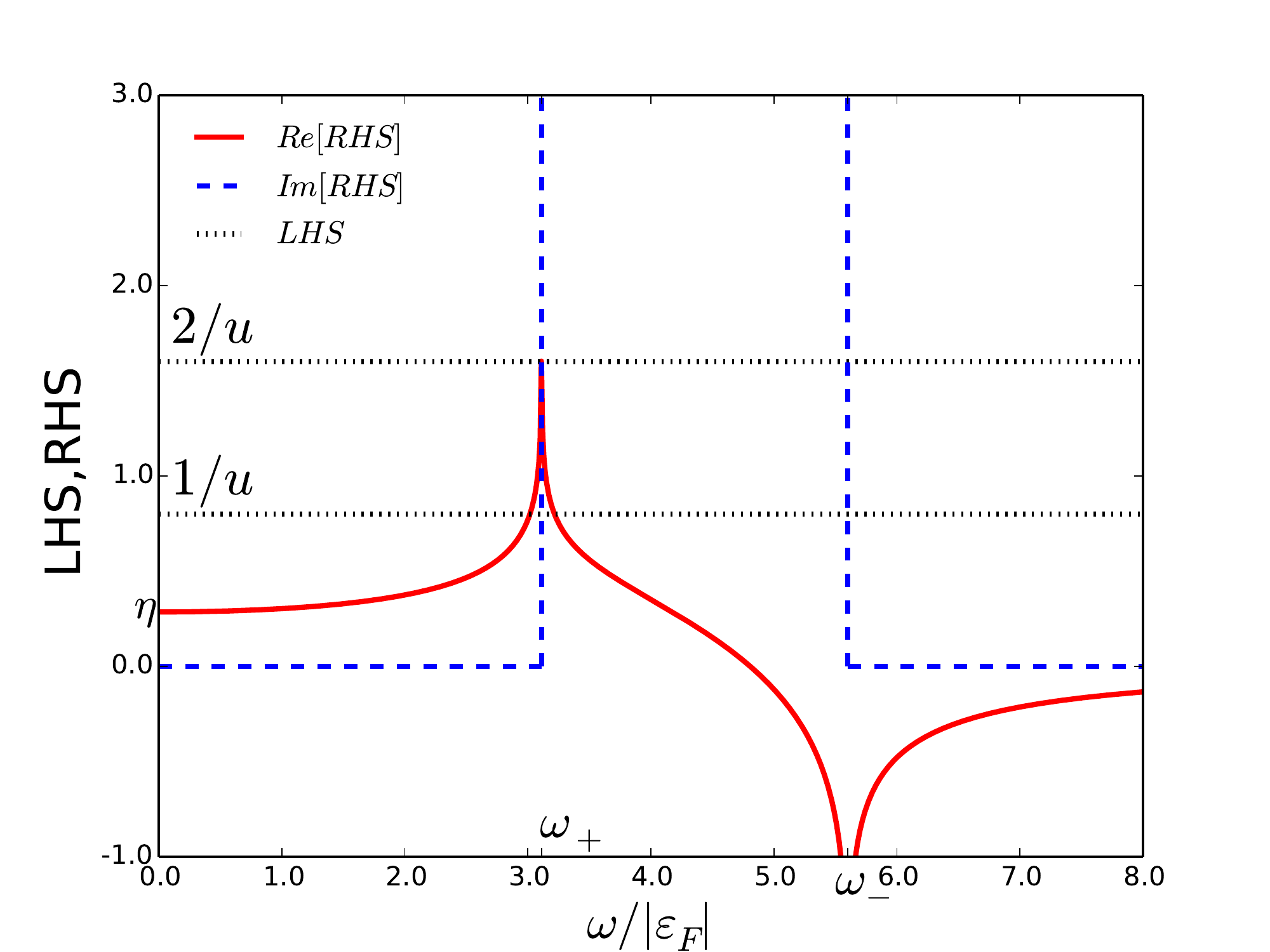}
\caption{Graphical solution of Eq.~\eqref{eq:w11_33} in one band regime  for $\lambda=1.75$ and $u=1.25$. 
Note that the real part of $L(\omega)$ has logarithmic singularities at $\omega_+$ and $\omega_-$ and its imaginary part is nonzero only for $\omega_+<\omega<\omega_-$, which corresponds to interband particle-hole excitation continuum at $q=0$. 
\label{fig:w11_w33}}
\end{figure}

For an arbitrary interaction strength $u$, Eqs.~\eqref{eq:w11_33} could be solved numerically. Following Ref.~\cite{Maiti2015}, we have plotted both sides of Eqs.~\eqref{eq:w11_33} in Fig.~\ref{fig:w11_w33}. 
The real part of $L(\omega)$ has logarithmic singularities at $\omega_\pm$, and its imaginary part is nonzero only for  $\omega_{+}<\omega<\omega_{-}$. This guarantees that for $u <1/\eta$ all three massive modes are undamped in the $q\to 0$ limit. For $1/\eta<u<2/\eta$, the perpendicular mode becomes damped and for larger interaction strengths all three modes will enter the interband particle-hole excitation continuum. As $\eta$ could be tuned through the Fermi energy in the single-band regime, therefore even for strongly interacting systems it is possible to find undamped massive modes at low enough particle densities. 

The remaining (i.e., the density) mode has an acoustic characteristic i.e., $\omega(q\to0)\propto q$, and its long wavelength behavior could be obtained from Eq.~\eqref{mode1_u}, after replacing $\chi^0_{0,0}$, $\chi^0_{0,2}$ and $\chi^0_{2,2}$ with their vanishing frequency and wave vector expansions [see, Eqs.~\eqref{eq:q-ac} in Appendix~\ref{app:rashba} for details] which simplifies to
\be\label{Zero_sound_one_band}
\begin{split}
\left[1-u I(y)\right]&
 \left[1-\frac{u}{2}(1+\eta)+ u y^2 
I(y)\right]
\\
&+\left[\frac{u y}{\sqrt{1+\bar{\varepsilon}_{F}/\lambda}}I(y)\right]^2=0,
\end{split}
\ee
where $y=\omega/(\alpha q \sqrt{1+\bar{\varepsilon}_{F}/\lambda})$ and
\be\label{eq:iy}
I(y)=\frac{\vert y \vert}{\sqrt{y^2-1}}\Theta(\vert y \vert -1)-1.
\ee
The numerical solution of this zero-sound mode in the two band regime is provided in Ref.~\cite{Zhang_pra2013}.

\subsection{Collective modes of a 2DEG with Rashba SOC}
\begin{figure}
      \includegraphics[width=0.5\textwidth]{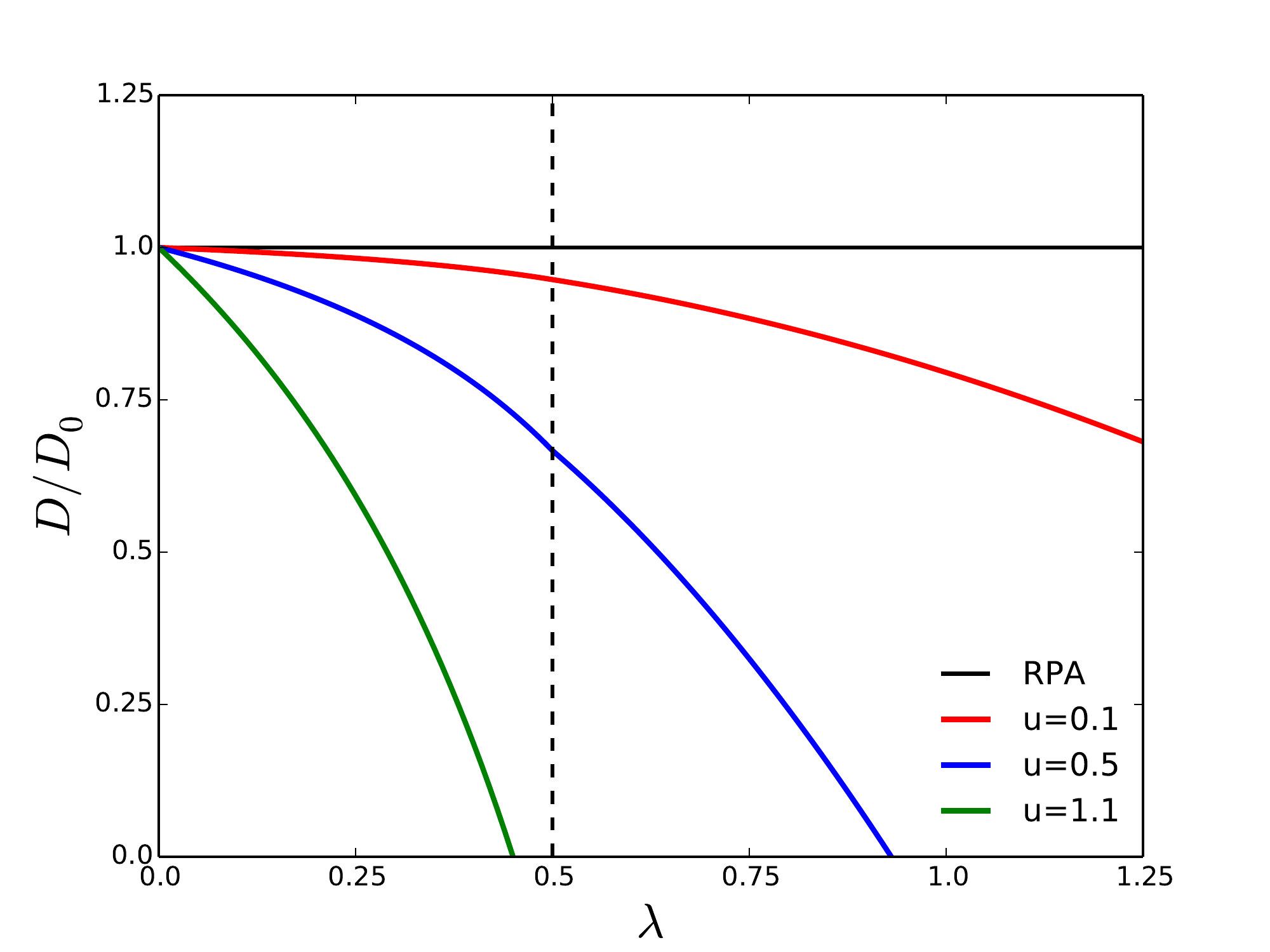}
    \caption{The Drude weight (in units of its noninteracting value $D_0$) is plotted versus the dimensionless SOC strength $\lambda$ for different values of the interaction strength $u$. The vertical dashed line at $\lambda=0.5$ indicates the border between two band regime at small SOC and single band regime at large SOC. \label{fig:drude}}
\end{figure}

In an electron gas system, the Fourier transform of the Coulomb interaction between electrons in 2D is $v(q)=2\pi e^2/q$ and the Hubbard potential $v_{\rm H}(q)$ approaches a constant value at long wavelengths. Similar to the ultracold systems with ultrashort interaction, solutions of Eqs.~\eqref{mode1} and~\eqref{mode2} for Rashba 2DEG will give three massive and one massless modes. The gapes of the massive modes would be identical to the ones of an ultracold system after replacing $U \rightarrow v_{\rm H}(q=0)$ in Eq.~\eqref{eq:w11_33}, however their full dispersions would be different due to the $q$-dependance of $v(q)$ and $v_{\rm H}(q)$. 
The massless mode, on the other hand, is a plasmon mode with $\omega(q\to 0) \propto \sqrt{q}$.
The strength of this plasmon mode at long wavelengths could be obtained from Eq.~\eqref{mode1}, after replacing the noninteracting response functions with their dynamical long wavelength (i.e., $q\to 0$, then $\omega\to 0$ limit) expansions [see, Eq.\eqref{eq:q-pl} in Appendix~\ref{app:rashba}], which up to leading order terms in $q$, reads
\be\label{eq:long-wavelength}
\begin{split}
&\left[1-v(q)\chi^{0}_{0,0}(q\to 0,\omega)\right]\left[1+\frac{v_{H}(0)}{2}\chi^{0}_{2,2}(q\to 0,\omega)\right]
\\
&+v(q)\frac{v_{H}(0)}{2}\left[\chi^{0}_{0,2}({q\to 0,\omega})\right]^2 =0.
\end{split}
\ee
Note that, as $v(q)$ diverges at $q\to 0$, its proper inclusion in Eq.~\eqref{eq:long-wavelength} is crucial and together with the coupling term $\chi^0_{0,2}$ would be responsible for giving the correct plasmon dispersion and the interaction induced modification of the Drude weight and plasmon mass.
After replacing the $q\to 0$ expansions of the non-interacting response functions in Eq.~\eqref{eq:long-wavelength}, we find
\be
\omega^2_{\rm pl}(q \to 0)= 2 D q +{\cal O}(q^2),
\ee
where $D$ is the Drude weight which is given by 
\begin{equation}\label{Drude}
 \frac{D}{D_{0}}= 1- \frac{u\lambda}{(2-{u}\eta)( \bar{\varepsilon}_F+\lambda)} .
\end{equation} 
Here, $u=m v_{\rm H}(q=0)/(2\pi)$, and $D_0=(n \pi e^2/m)(\bar{\varepsilon}_F+\lambda)$ is the non-interacting Drude weight of a 2D Rashba system. In the single band regime Eq.~\eqref{Drude} simplifies to
\begin{equation}\label{drude_1band}
 \frac{D}{D_{0}}= 1- \frac{4u\lambda^2}{2-{u/(2\lambda)}}, ~~~~\textrm{for}~\varepsilon_F<0,
\end{equation} 
 while in the two band regime we find
\begin{equation}\label{drude_2band}
 \frac{D}{D_{0}}= 1- \frac{u\lambda}{(2-{u})( 1-\lambda)}, ~~~~\textrm{for}~\varepsilon_F>0.
\end{equation} 
In Fig.~\ref{fig:drude} we have illustrated the suppression of Drude weight due to spin-orbit coupling and interparticle interaction for different values of the interaction strength. 
Similar results in the two-band regime for the plasmon mass from the time-dependent Hartree-Fock method has been obtained by Agarwal \textit{et al.}~\cite{Amit2011} and for the optical conductivity from diagrammatic calculations by Maiti \textit{et al.}~\cite{Maiti2015}. However, this enhancement of the plasmon mass has been entirely dismissed in the diagrammatic approach of  Ref.~\cite{Maiti2015}.
We would like to point out that Eq.~\eqref{Drude} also predicts the enhancement of the Drude weight for $u>2/\eta$, but our results in this regime of very strong interaction should be interpreted with caution.


\subsection{Spin-Hall conductivity}
The spin-Hall conductivity measures the $z$-component of the spin-current flowing in the direction transverse to the applied electric field. If we assume that our system is subjected to a uniform (i.e., q=0) electric field in the $x$ direction, using the equation of motion method~\cite{Amit2011} it could be shown that the spin-Hall conductivity for a Rashba system is given by
\be\label{Optical-Hall}
\sigma_{SH}(\omega)=\frac{e}{4m }\chi_{2,2}(q=0,\omega),
\ee
where $\chi_{2,2}$ represents a component of the interacting spin-density response function, introduced through Eq.~\eqref{eq:s_chi_phi}. 
For a non-interacting system, replacing the $q=0$ limit of $\chi^0_{2,2}$ from Eq.~\eqref{eq:q=0}, in the static limit we find
\be\label{non-interacting-optical-Hall}
\sigma^{0}_{SH}(\omega=0)=\frac{e}{4m}(-\eta\frac{m}{2\pi})=-\frac{e}{8\pi} \eta,
\ee
which in the two-band regime (i.e., $\eta=1$) reduces to the famous universal value $-e/(8\pi)$ for the intrinsic spin-Hall conductivity of Rashba system~\cite{JSinova2004}.
For an interacting system, it could be easily shown from Eq.~\eqref{eq:s_chi_phi} that 
\be\label{Interacting-chi22}
\chi_{2,2}(0,\omega)=\frac{\chi^{0}_{2,2}(0,\omega)}{1+v_{\rm H}(0)\chi^{0}_{2,2}(0,\omega)/2}.
\ee
This gives the interaction induced correction to the spin-Hall conductivity as
\be\label{eq:spin-Hall} 
\sigma_{SH}(\omega=0)=\frac{\sigma^{0}_{SH}(\omega=0)}{1-\eta u/2}, 
\ee 
which in the two-band regime is dientical to the results obtained by Agarwal \textit{et al.}~\cite{Amit2011} in the ultrashort-range interaction limit. 

\begin{figure}
\centering
\includegraphics[width=0.5\textwidth]{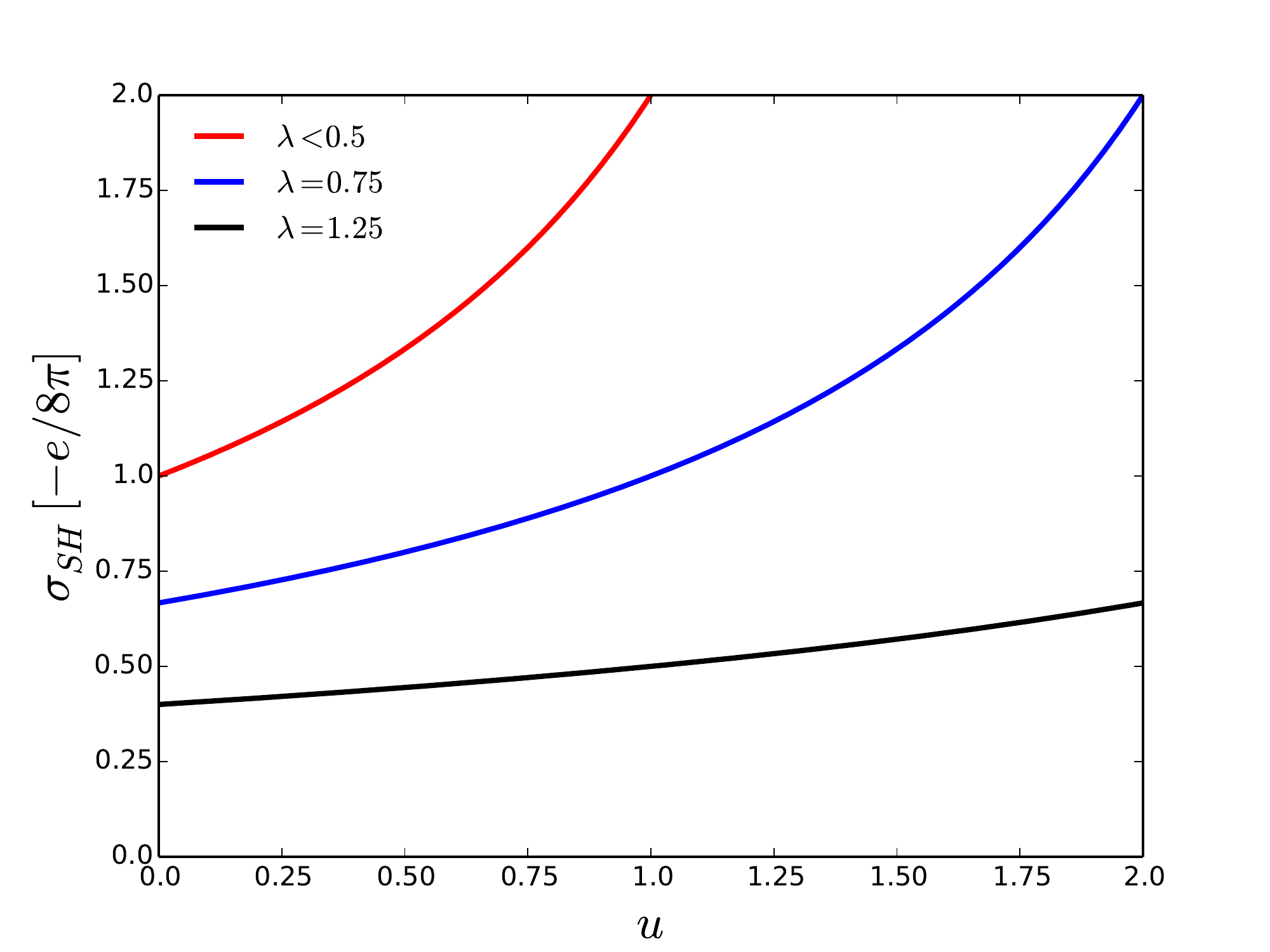}
\caption{The intrinsic contribution to the optical spin-Hall conductivity of 2D Rashba system [in units of $-e/(8\pi)$] versus the dimensionless interaction strength $u$. 
The red solid curve shows the results in the two band regime i.e., $\lambda<0.5$ where the spin-Hall conductivity is independent of the SOC strength. Blue and black curves are the results for two different values of the SOC in the single band regime.}
\label{fig:Spin-Hall}
\end{figure}

It is interesting to notice that the correction to the universal value of spin-Hall conductivity in the two-band regime depends only on the interaction strength, but in the single-band regime this correction is a function of interaction strength, SOC strength and the particle density (see, Fig.~\ref{fig:Spin-Hall}).
Finally, we should also mention that the result in Eq.~\eqref{eq:spin-Hall} is applicable to both ultracold and electronic systems with Rashba SOC.

\section{Summary and discussion}\label{sect:summary}

In summary, we have used the equation of motion for single particle distribution function of a Fermi liquid with spin-orbit coupling. The generated higher order distribution functions have been reduced to single particle ones with a Hartree-Fock mean field approximation and the nonlocal exchange term has been approximated with a local one. This lead to the introduction of a Hubbard like many-body local field factor for helical systems and would in principle give the dispersions of four coupled spin-charge collective modes. Application of this simple formalism to a two-dimensional Rashba system proves its power in providing reasonable results for several physical quantities.

We should note that the present formulation should be considered as the first step towards introducing the concept of local field factors for helical systems and much more improvements are yet to be made to find the general forms of the local field factors and their asymptotic and/or dynamical behaviors. In particular, we expect that in their most general form, the many-body local field factors would depend on the spin indices and would also induce off-diagonal components in the effective interaction matrix of Eq. \eqref{eq:effective-interaction}.


\acknowledgments
The authors gratefully acknowledge support by the Institute for Advanced Studies in Basic Sciences (IASBS) Research Council under grant No. G2016IASBS12646.
We also thank the kind hospitality of condensed matter and statistical physics section of ICTP, Trieste during the final stages of the preparation of this work.

\appendix

\begin{widetext}

\section{Details of the equation of motion method}\label{app:eom}

In order to obtain the spin-density linear response functions, we write the Heisenberg equation of motion for the single particle distribution function
$\hdg{c}_{\kv-\qv,\mu} \h{c}_{\kv,\nu} $~\cite{solyom_v3,niklasson_prb74}
\be
\begin{split}
i \frac{d}{dt} \hdg{c}_{\kv-\qv,\mu} \h{c}_{\kv,\nu}
= \left[\hdg{c}_{\kv-\qv,\mu} \h{c}_{\kv,\nu},\h{H}\right].
\end{split}
\ee
The Hamiltonian $\h{H}$ of an interacting Fermi system with a generic spin-orbit coupling (SOC) and a spin-dependent external potential is given in Eq. \eqref{hamil}, so we have
\be\label{eom2}
\begin{split}
i \frac{d}{dt} \hdg{c}_{\kv-\qv,\mu} \h{c}_{\kv,\nu}
=  \left[\hdg{c}_{\kv-\qv,\mu} \h{c}_{\kv,\nu},\h{T}\right]+ \left[\hdg{c}_{\kv-\qv,\mu} \h{c}_{\kv,\nu},\h{V}_{\rm ext}\right]
+ \left[\hdg{c}_{\kv-\qv,\mu} \h{c}_{\kv,\nu},\h{V}_{\rm int}\right].
\end{split}
\ee
The evaluation of first term on the RHS of Eq. \eqref{eom2} is very straightforward and gives Eq. \eqref{eq:cc-T}. The  second term on the RHS of  Eq. \eqref{eom2} results in
\be\label{cc-Vext_0}
\begin{split}
\left[\hdg{c}_{\kv-\qv,\mu} \h{c}_{\kv,\nu},\h{V}_{\rm ext}\right]&=
\frac{1}{\Vc}\sum_{j,\qv',\gamma}\phi^j_{\rm ext}(-\qv')
\left[F^j_{\nu,\gamma}(\kv;\kv+\qv') \hdg{c}_{\kv-\qv,\mu} \h{c}_{\kv+\qv',\gamma}
-F^i_{\gamma,\mu}(\kv-\qv-\qv';\kv-\qv) \hdg{c}_{\kv-\qv-\qv',\gamma} \h{c}_{\kv,\nu}\right],
\end{split}
\ee
which after taking the expectation values of its both sides, and discarding the nonlinear terms in the external potential, gives Eq. \eqref{eq:cc-Vext}.
The evaluation of third term on the RHS of Eq. \eqref{eom2} is more involved and below we provide its  main steps in details
\be\label{cc-Vint_1}
\begin{split}
\left[\hdg{c}_{\kv-\qv,\mu} \h{c}_{\kv,\nu},\hat{V}_{int}\right]&=\frac{1}{2\Vc} \sum_{\kv',\mu',\nu'} \sum_{\kv'',\mu'',\nu''}\sum_{\qv'} v(\qv') F^{0}_{\mu',\nu'}(\kv'-\qv',\kv') F^{0}_{\mu'',\nu''}(\kv''+\qv',\kv'')\\
&\times\left[\hdg{c}_{\kv-\qv,\mu} \h{c}_{\kv,\nu},\hdg{c}_{\kv'-\qv',\mu'} \h{c}_{\kv',\nu'} \hdg{c}_{\kv''+\qv',\mu''} \h{c}_{\kv'',\nu''} \right].
\end{split}
\ee
The commutator of creation and annihilation operators on the RHS of Eq. \eqref{cc-Vint_1} gives  
\be\label{bracket-1}
\begin{split}
[\hdg{c}_{\kv-\qv,\mu} \h{c}_{\kv,\nu},\hdg{c}_{\kv'-\qv',\mu'} &\hdg{c}_{\kv''+\qv',\mu''}\h{c}_{\kv'',\nu''} \h{c}_{\kv',\nu'}]\\
&=\delta_{\kv,\kv'-\qv'}\delta_{\nu,\mu'} \hdg{c}_{\kv-\qv,\mu} \hdg{c}_{\kv''+\qv',\mu''} \h{c}_{\kv'',\nu''} \h{c}_{\kv',\nu'}-\delta_{\kv,\kv''+\qv'}\delta_{\nu,\mu''} \hdg{c}_{\kv-\qv,\mu} \hdg{c}_{\kv'-\qv',\mu'} \h{c}_{\kv'',\nu''} \h{c}_{\kv',\nu'}\\
&+\delta_{\kv'',\kv-\qv}\delta_{\nu'',\mu} \hdg{c}_{\kv'-\qv',\mu'} \hdg{c}_{\kv''+\qv',\mu''} \h{c}_{\kv',\nu'} \h{c}_{\kv,\nu}-\delta_{\kv',\kv-\qv}\delta_{\nu',\mu} \hdg{c}_{\kv'-\qv',\mu'} \hdg{c}_{\kv''+\qv',\mu''} \h{c}_{\kv'',\nu''} \h{c}_{\kv,\nu}.
\end{split}
\ee
After replacing Eq. \eqref{bracket-1} in Eq. \eqref{cc-Vint_1}, we find
\be\label{cc-Vint_2}
\begin{split}
\left[\hdg{c}_{\kv-\qv,\mu} \h{c}_{\kv,\nu},\hat{V}_{int}\right]&=\frac{1}{2\Vc}\sum_{\qv',\nu'}\sum_{\kv'',\mu'',\nu''} v(\qv') F^{0}_{\nu,\nu'}(\kv,\kv+\qv') F^{0}_{\mu'',\nu''}(\kv''+\qv',\kv'') \hdg{c}_{\kv-\qv,\mu} \hdg{c}_{\kv''+\qv',\mu''} \h{c}_{\kv'',\nu''} \h{c}_{\kv+\qv',\nu'}
 \\
&-\frac{1}{2\Vc}\sum_{\qv',\nu''}\sum_{\kv',\mu',\nu'} v(\qv')F^{0}_{\mu',\nu'}(\kv'-\qv',\kv') F^{0}_{\nu,\nu''}(\kv,\kv-\qv') \hdg{c}_{\kv-\qv,\mu} \hdg{c}_{\kv'-\qv',\mu'} \h{c}_{\kv-\qv',\nu''} \h{c}_{\kv',\nu'}
 \\
&+\frac{1}{2\Vc}\sum_{\qv',\mu''}\sum_{\kv',\mu',\nu'} v(\qv')F^{0}_{\mu',\nu'}(\kv'-\qv',\kv') F^{0}_{\mu'',\mu}(\kv-\qv+\qv',\kv-\qv) \hdg{c}_{\kv'-\qv',\mu'} \hdg{c}_{\kv-\qv+\qv',\mu''} \h{c}_{\kv',\nu'} \h{c}_{\kv,\nu}
\\
&-\frac{1}{2\Vc}\sum_{\qv',\mu'}\sum_{\kv'',\mu'',\nu''} v(\qv')F^{0}_{\mu',\mu}(\kv-\qv-\qv',\kv-\qv) F^{0}_{\mu'',\nu''}(\kv''+\qv',\kv'') \hdg{c}_{\kv-\qv-\qv',\mu'} \hdg{c}_{\kv''+\qv',\mu''} \h{c}_{\kv'',\nu''} \h{c}_{\kv,\nu}~.
\end{split}
\ee
If we replace  $\kv''\longrightarrow \kv'$, $\qv'\longrightarrow -\qv'$, $\mu''\longrightarrow \mu'$, and $\nu'' \longleftrightarrow \nu'$  in the first line on the RHS of Eq. \eqref{cc-Vint_2}, and $\kv''\longrightarrow \kv'$, $\qv'\longrightarrow -\qv'$, $\mu''\longleftrightarrow \mu'$, and $\nu'' \longrightarrow \nu'$ in its last line, after some rearrangements we obtain
\be\label{cc-Vint_3}
\begin{split}
&\left[\hdg{c}_{\kv-\qv,\mu} \h{c}_{\kv,\nu},\hat{V}_{int}\right]
=\frac{1}{\Vc} \sum_{\qv'}v(\qv')\sum_{\kv',\mu',\nu'}
F^{0}_{\mu',\nu'}(\kv'-\qv';\kv')\\
&\times\sum_\gamma\left[
F^{0}_{\nu,\gamma}(\kv;\kv-\qv')\hdg{c}_{\kv-\qv,\mu} \hdg{c}_{\kv'-\qv',\mu'}\h{c}_{\kv',\nu'}\h{c}_{\kv-\qv',\gamma}-
F^{0}_{\gamma,\mu}(\kv-\qv+\qv';\kv-\qv)\hdg{c}_{\kv-\qv+\qv',\gamma} \hdg{c}_{\kv'-\qv',\mu'}\h{c}_{\kv',\nu'}\h{c}_{\kv,\nu}\right].
\end{split}
\ee
Now, using the mean-field decoupling of the quartic terms into quadratic terms in creation and annihilation operators
\be
\hdg{c}_1\hdg{c}_2\h{c}_3\h{c}_4 \approx  \langle \hdg{c}_1\h{c}_4\rangle  \hdg{c}_2 \h{c}_3
+ \langle \hdg{c}_2\h{c}_3\rangle  \hdg{c}_1 \h{c}_4
- \langle \hdg{c}_1\h{c}_3\rangle  \hdg{c}_2 \h{c}_4
- \langle \hdg{c}_2\h{c}_4\rangle  \hdg{c}_1 \h{c}_3,
\ee
in the first line on the RHS of Eq. \eqref{cc-Vint_3} we find
\be\label{decoupling}
\begin{split}
\hdg{c}_{\kv-\qv,\mu} \hdg{c}_{\kv'-\qv',\mu'}\h{c}_{\kv',\nu'}\h{c}_{\kv-\qv',\gamma}\approx &\langle \hdg{c}_{\kv-\qv,\mu} \h{c}_{\kv-\qv',\gamma}\rangle \hdg{c}_{\kv'-\qv',\mu'}\h{c}_{\kv',\nu'}+\langle \hdg{c}_{\kv'-\qv',\mu'}\h{c}_{\kv',\nu'}\rangle  \hdg{c}_{\kv-\qv,\mu}\h{c}_{\kv-\qv',\gamma}\\
-&\langle \hdg{c}_{\kv-\qv,\mu} \h{c}_{\kv',\nu'}\rangle \hdg{c}_{\kv'-\qv',\mu'}\h{c}_{\kv-\qv',\gamma}-\langle \hdg{c}_{\kv'-\qv',\mu'}\h{c}_{\kv-\qv',\gamma}\rangle  \hdg{c}_{\kv-\qv,\mu}\h{c}_{\kv',\nu'}.
\end{split}
\ee
To linear order in external perturbations, the expectation values in Eq.~\eqref{decoupling} could be evaluated with respect to the equilibrium state. It immediately becomes clear that the second and fourth terms on the RHS of  Eq.~\eqref{decoupling} correspond to the self-energy corrections to the single particle energies within the Hartree-Fock approximation and in principle their diagonal (intra-band) part could be easily absorbed into Eq.~\eqref{eq:cc-T} with a redefinition of the single particle energies. 
We will ignore these self energy terms throughout this paper as our discussion is not based on the specific form of the band structure. With a similar approximation for the second term on the RHS of Eq.~\eqref{cc-Vint_3}, and after some straightforward change of variables, we find
\be
\begin{split}
\left[\hdg{c}_{\kv-\qv,\mu} \h{c}_{\kv,\nu},\hat{V}_{int}\right]
&\approx\frac{1}{\Vc}\sum_{\kv',\mu',\nu'}v(\qv)F^{0}_{\mu',\nu'}(\kv'-\qv;\kv')F^{0}_{\nu,\mu}(\kv;\kv-\qv)\hdg{c}_{\kv'-\qv,\mu'}\h{c}_{\kv',\nu'}n_{\kv-\qv,\mu}\\
&-\frac{1}{\Vc}\sum_{\kv',\mu',\nu'}v(\kv-\kv')F^{0}_{\mu',\mu}(\kv'-\qv;\kv-\qv)F^{0}_{\nu,\nu'}(\kv;\kv')\hdg{c}_{\kv'-\qv,\mu'}\h{c}_{\kv',\nu'}n_{\kv-\qv,\mu}\\
&-\frac{1}{\Vc}\sum_{\kv',\mu',\nu'} v(\qv) F^{0}_{\mu',\nu'}(\kv'-\qv;\kv')F^{0}_{\nu,\mu}(\kv;\kv-\qv) \hdg{c}_{\kv'-\qv,\mu'}\h{c}_{\kv',\nu'}n_{\kv,\nu}\\
&+\frac{1}{\Vc}\sum_{\kv',\mu',\nu'}  v(\kv'-\kv) F^{0}_{\nu,\nu'}(\kv;\kv')F^{0}_{\mu',\mu}(\kv'-\qv;\kv-\qv)  \hdg{c}_{\kv'-\qv,\mu'}\h{c}_{\kv',\nu'}n_{\kv,\nu},
\end{split}
\ee
which after some rearrangements results in  
\be\label{cc-Vint_4}
\begin{split}
\left[\hdg{c}_{\kv-\qv,\mu} \h{c}_{\kv,\nu},\hat{V}_{int}\right]&\approx
\frac{1}{\Vc}\left(n_{\kv-\qv,\mu}-n_{\kv,\nu}\right) v(q) F^{0}_{\nu,\mu}(\kv;\kv-\qv)
 \sum_{\kv',\mu',\nu'}F^{0}_{\mu',\nu'}(\kv'-\qv;\kv') \hdg{c}_{\kv'-\qv,\mu'} \h{c}_{\kv',\nu'}  \\
&-\frac{1}{\Vc}\left(n_{\kv-\qv,\mu}-n_{\kv,\nu}\right) \sum_{\kv',\mu',\nu'} v(\kv-\kv') F^{0}_{\nu,\nu'}(\kv;\kv') F^{0}_{\mu',\mu}(\kv'-\qv;\kv-\qv) \hdg{c}_{\kv'-\qv,\mu'} \h{c}_{\kv',\nu'}.
\end{split}
\ee
Now, if we take the expectation values of both side of Eq. \eqref{cc-Vint_4} we arrive at Eq. \eqref{eq:cc-P}.

\end{widetext}

\section{Noninteracting spin-density response functions of Rashba 2DEG}\label{app:rashba}
The elements of linear spin-density response matrix for noninteracting 2D electron gas with Rashba spin-orbit coupling as defined in Eq.~(\ref{eq:chi0_matrix}) is
\be
\begin{split}
&\chi^{0}_{i,j}(\qv,\omega)=\langle\langle  S^{i}_{\qv}; S^{j}_{-\qv}\rangle\rangle_\omega \\
&=\frac{1}{{\cal S}}\sum_{\kv,\mu,\nu} 
\frac{n_{\kv-\qv,\mu}-n_{\kv,\nu}}{\omega +\varepsilon_{\kv-\qv,\mu}-\varepsilon_{\kv,\nu}}
F^{i}_{\mu,\nu}(\kv-\qv;\kv)F^{j}_{\nu,\mu}(\kv;\kv-\qv),
\end{split}
\ee
where ${\cal S}$ is the sample area and the form factors are given in Eq.~(\ref{eq:formfactors}). In order to calculate different elements of the linear spin-density response matrix we need different combinations of the form factors $F^{i}_{\mu,\nu}(\kv-\qv;\kv)F^{j}_{\nu,\mu}(\kv;\kv-\qv)$, which are simply obtained as  
\be\label{eq:product_of_formfactors}
\begin{split}
&F^{0}F^{0}=\frac{1}{2}\left[1+\mu \nu \cos(\varphi_{k-q}-\varphi_{k}) \right],\\
&F^{0}F^{1}=-\frac{1}{2}\left[\mu \sin(\varphi_{k-q})+\nu \sin(\varphi_{k})\right],\\
&F^{0}F^{2}=\frac{1}{2}\left[\mu \cos(\varphi_{k-q})+ \nu \cos(\varphi_{k}) \right],\\
&F^{0}F^{3}=-\frac{i}{2}\mu \nu \sin(\varphi_{k-q}-\varphi_{k}),\\
&F^{1}F^{0}=F^{0}F^{1},\\
&F^{1}F^{1}=\frac{1}{2}\left[1-\mu \nu \cos(\varphi_{k-q}+\varphi_{k}) \right],\\
&F^{1}F^{2}=-\frac{1}{2}\mu \nu \sin(\varphi_{k-q}+\varphi_{k}),\\
&F^{1}F^{3}=-\frac{i}{2}\left[\mu \cos(\varphi_{k-q})- \nu \cos(\varphi_{k}) \right],\\
&F^{2}F^{0}=F^{0}F^{2},\\
&F^{2}F^{1}=F^{1}F^{2},\\
&F^{2}F^{2}=\frac{1}{2}\left[1+\mu \nu \cos(\varphi_{k-q}+\varphi_{k}) \right],\\
&F^{2}F^{3}=-\frac{i}{2}\left[\mu \sin(\varphi_{k-q})-\nu \sin(\varphi_{k})\right],\\
&F^{3}F^{0}=-F^{0}F^{3},\\
&F^{3}F^{1}=-F^{1}F^{3},\\
&F^{3}F^{2}=-F^{2}F^{3},\\
&F^{3}F^{3}=\frac{1}{2}\left[1-\mu \nu \cos(\varphi_{k-q}-\varphi_{k}) \right],
\end{split}
\ee
where the arguments and indices are omitted on the left hand sides for brevity. 
 
The only nonzero elements of the linear spin-density response function matrix are $\chi^{0}_{0,0}$, $\chi^{0}_{1,1}$, $\chi^{0}_{2,2}$, $\chi^{0}_{3,3}$, $\chi^{0}_{0,2}=\chi^{0}_{2,0}$ and $\chi^{0}_{1,3}=-\chi^{0}_{3,1}$~\cite{Maiti2015}. Below we will discuss the long wavelength behaviors of these responses. For detailed results please refer to Ref.~\cite{Maiti2015}. 

At finite $\omega$, and $q=0$, all elements of the density response matrix for Rashba system vanish except $\chi^{0}_{1,1}$, $\chi^{0}_{2,2}$ and $\chi^{0}_{3,3}$, which read
\be\label{eq:q=0}
\begin{split}
\chi^{0}_{1,1}(q= 0,\omega)&=
\chi^{0}_{2,2}(q= 0,\omega)=
\chi^{0}_{3,3}(q= 0,\omega)/2
\\
&=-\nu_0\left[\eta+\frac{\omega}{8 m \alpha^2}L(\omega)\right],
\end{split}
\ee
Here, $\nu_0=m/(2\pi)$ and $\eta$ and $L(\omega)$ are defined in the main text after Eq.~\eqref{eq:w11_33}.

Taking $q \to 0$ and then $\omega \to 0$ limit, which is relevant for the dispersion of plasmon mode in the long wavelength limit, the leading order terms of the relevant noninteracting spin-density response functions read
\begin{equation}\label{eq:q-pl}
\begin{split}
\chi^{0}_{0,0}(q\to 0,\omega\to 0)&=\frac{n}{m}\left(\lambda+\bar{\varepsilon}_{F}\right) \left(\frac{q}{\omega}\right)^2,\\
\chi^{0}_{2,2}(q\to 0,\omega\to 0)&=-\nu_0\eta,\\
\chi^{0}_{0,2}(q\to 0,\omega \to 0)&=-\nu_0\alpha \left(\frac{q}{\omega}\right).
\end{split}
\end{equation} 
On the other hand, if the $q\to 0$ and $\omega \to 0$ limits are taken simultaneously, such that the dimensionless ratio $y=\omega/(\alpha q \sqrt{1+\bar{\varepsilon}_{F}/\lambda})$ is kept fixed, we obtain
\begin{equation}\label{eq:q-ac}
\begin{split}
\chi^{0}_{0,0}(q\to 0,y)&=2 \nu_0I(y), \\
\chi^{0}_{2,2}(q\to 0,y)&=-\nu_0(1+\eta)+y^2\chi^{0}_{0,0}(q\to 0,y),\\
\chi^{0}_{0,2}(q\to 0,y)&=-\frac{y }{\sqrt{1+\bar{\varepsilon}_{F}/\lambda}}\chi^{0}_{0,0}(q\to 0,y),
\end{split}
\end{equation} 
where $I(y)$ is defined in Eq.~\eqref{eq:iy}.



\begin{thebibliography}{90}
\bibitem{Awshalom2009}
D. Awschalom and N. Samarth, Physics \textbf{2}, 50 (2009).
%
\bibitem{Zutic}
I. \v Zuti\' c, J. Fabian, and S. Das Sarma, Rev. Mod. Phys. \textbf{76}, 323~(2004).
%
\bibitem{Manchon2015}
A. Manchon, H. C. Koo, J. Nitta, S. M. Frolov, and R. A. Duine, Nat. Mater \textbf{14}, 871–882 (2015).
%
\bibitem{ZHassan2010}
M. Z. Hasan and C. L. Kane, Rev. Mod. Phys. \textbf{82}, 3045 (2010).
%
\bibitem{LiangFu2008}
L. Fu and C. L. Kane, Phys. Rev. Lett. \textbf{100}, 096407 (2008). 
%
\bibitem{dalibard_rmp2011} 
 J. Dalibard, F. Gerbier, G. Juzeli\={u}nas, and P. \"{O}hberg, Rev. Mod. Phys. \textbf{83}, 1523 (2011).
%
\bibitem{goldman_rpp2014}
 N. Goldman, G. Juzeli\={u}nas, P. \"{O}hberg, I. B. Spielman, Rep. Prog. Phys. \textbf{77}, 126401 (2014).
%
\bibitem{YPlotnik2016}
  Y. Plotnik, M. A. Bandres, S. St{\"u}tzer, Y. Lumer, M. C. Rechtsman, A. Szameit, and M. Segev, Phys. Rev. B \textbf{94}, 020301(R) (2016).
%
\bibitem{abbaszadeh_arxiv2016}
  H. Abbaszadeh, A. Souslov, J. Paulose, H. Schomerus, and V. Vitelli, arXiv:1610.06406 (unpublished).
%
\bibitem{Ghchen1999-1}
 G.-H. Chen and M. E. Raikh, Phys. Rev. B \textbf{59}, 5090 (1999).
%
\bibitem{Ghchen1999-2}
 G.-H. Chen and M. E.Raikh, Phys. Rev. B \textbf{60}, 4826 (1999).
%
\bibitem{Magarill2001}
 L. I. Magarill, A. V. Chaplik, and M. V. \'Entin, JETP \textbf{92}, 153 (2001).
%
\bibitem{Saraga2005} 
 D. S. Saraga and D. Loss, Phys. Rev. B \textbf{72}, 195319 (2005).
%
\bibitem{Dimitrova2005} 
  O. V. Dimitrova, Phys. Rev. B \textbf{71}, 245327 (2005).
%
\bibitem{Wang2005} 
 X. F. Wang, Phys. Rev. B \textbf{72}, 085317 (2005). 
 %
\bibitem{Pletyukhov2006} 
 M. Pletyukhov and V. Gritsev, Phys. Rev. B \textbf{74}, 045307 (2006). 
%
\bibitem{Schliemann2006} 
 J. Schliemann, Phys. Rev. B \textbf{74}, 045214 (2006). 
%
 \bibitem{Chesi2007} 
 S. Chesi and G. F. Giuliani, Phys. Rev. B \textbf{75}, 153306 (2007).
%
\bibitem{Badalyan2009}
 S. M. Badalyan, A. Matos-Abiague, G. Vignale, and J. Fabian, Phys. Rev. B \textbf{79}, 205305 (2009).
%
\bibitem{Ambrosetti2009}
 A. Ambrosetti, F. Pederiva, E. Lipparini, and S. Gandolfi, Phys. Rev. B \textbf{80}, 125306 (2009).
%
\bibitem{Nechaev2010}
 I. A. Nechaev, P. M. Echenique, and E. V. Chulkov, Phys. Rev. B \textbf{81}, 195112 (2010).
%
\bibitem{Badalyan2010}
 S. M. Badalyan, A. Matos-Abiague, G. Vignale, and J. Fabian, Phys. Rev. B \textbf{81}, 205314 (2010).
%
\bibitem{Zak2010}
 R. A. \.Zak, D. L. Maslov, and D. Loss, Phys. Rev. B \textbf{82}, 115415 (2010).
%
\bibitem{Chesi2011-1}
S. Chesi and G. F. Giuliani, Phys. Rev. B \textbf{83}, 235308 (2011).
%
\bibitem{Chesi2011-2}
S. Chesi and G. F. Giuliani, Phys. Rev. B \textbf{83}, 235309 (2011).
%
\bibitem{Ashrafi2012} 
 A. Ashrafi and D. L. Maslov, Phys. Rev. Lett. \textbf{109}, 227201 (2012). 
%
\bibitem{Maiti2015-2}
 S. Maiti, M. Imran, and D. L. Maslov, Phys. Rev. B \textbf{93}, 045134 (2016).
%
\bibitem{Shekhter}
A. Shekhter, M. Khodas, and A. M. Finkel'stein, Phys. Rev. B \textbf{71}, 165329 (2005).
%
\bibitem{Farid2006}
A.-K. Farid and E. G. Mishchenko, Phys. Rev. Lett. \textbf{97}, 096604 (2006).
%
\bibitem{Amit2011}
A. Agarwal, S. Chesi, T. Jungwirth, J. Sinova, G. Vignale, and M. Polini, Phys. Rev. B \textbf{83}, 115135 (2011).
%
\bibitem{Maiti2015}
S. Maiti, V. Zyuzin, and D. L. Maslov, Phys. Rev. B \textbf{91}, 035106 (2015).
%
\bibitem{Abedinpour2011}
S. H. Abedinpour, G. Vignale, A. Principi, M. Polini, W.-K. Tse, and A. H.  MacDonald, Phys. Rev. B \textbf{84}, 045429 (2011).
%
\bibitem{GV_book}
 G. F. Giuliani and G. Vignale, \textit{Quantum Theory of the Electron Liquid} (Cambridge University Press, Cambridge, 2005).
%
\bibitem{Hubbard_1957}
 J. Hubbard Proc. Roy. Soc. (London) A \textbf{243}, 336 (1957). 
%
\bibitem{niklasson_prb74}
 G. Niklasson, Phys. Rev. B \textbf{10}, 3052 (1974).
%
\bibitem{solyom_v3}
 J. S\'{o}lyom, \textit{Fundamentals of the Physics of Solids, Volume 3 - Normal, Broken-Symmetry, and Correlated Systems} (Berlin, Springer-Verlag, 2010).
%
\bibitem{mahan_book}
 G. Mahan, \textit{Many-Particle Physics} (Klauwer/Plenium, 3rd ed., 2000).
%
\bibitem{Simion_thesis} 
 G. Simion, Ph.D thesis, Purdue University (2007).
%
\bibitem{foot:invariance}
 In performing the equation of motion the only assumption we have made is that the summations over wave vectors are unrestricted and one can easily shift the variables in the summation without any trouble. Therefore our formalism is not directly applicable \textit{e.g.}, for the effective low-energy models of graphene where an ultraviolet cutoff on momentum is necessary. 
%
\bibitem{Zhang_pra2013} 
 S.-S. Zhang, X.-L. Yu, J. Ye, and W.-M Liu, Phys Rev. A \textbf{87}, 063623 (2013).  
%
\bibitem{chin_rmp2010}
 C. Chin, R. Grimm, P. Julienne, and E. Tiesinga,  Rev. Mod. Phys. \textbf{82}, 1225 (2010).
%
\bibitem{JSinova2004}
 J. Sinova, D. Culcer, Q. Niu, N. A. Sinitsyn, T. Jungwirth, and A. H. MacDonald, Phys. Rev. Lett. \textbf{92}, 126603 (2004). 
%
\end{thebibliography}
\end {document}